\documentclass{article}

\usepackage[utf8]{inputenc}
\usepackage[T1]{fontenc}
\DeclareUnicodeCharacter{2032}{\textquotesingle}
\DeclareUnicodeCharacter{00B1}{+/-}
\DeclareUnicodeCharacter{2013}{--}
\DeclareUnicodeCharacter{2014}{---}
\DeclareUnicodeCharacter{2019}{'}
\DeclareUnicodeCharacter{201C}{``}
\DeclareUnicodeCharacter{201D}{''}
\DeclareUnicodeCharacter{2022}{*}
\DeclareUnicodeCharacter{2192}{\ensuremath{\rightarrow}}

\usepackage{microtype}
\usepackage{graphicx}
\usepackage{booktabs}

\usepackage{xurl}
\usepackage{hyperref}
\hypersetup{breaklinks=true}
\usepackage[preprint]{icml2026}

\usepackage{mdframed}
\usepackage{fvextra}
\usepackage{comment}
\usepackage{tabularx}
\usepackage{array,makecell}

\usepackage{amsmath,amssymb}

\icmltitlerunning{SAE-RNA: Interpreting RNA Language Model Representations}

\begin{document}

\twocolumn[
\icmltitle{SAE-RNA: A Sparse Autoencoder Model for Interpreting RNA Language Model Representations}

\icmlsetsymbol{equal}{*}

\begin{icmlauthorlist}
\icmlauthor{Taehan Kim}{equal,berkeleycs}
\icmlauthor{Sangdae Nam}{equal,berkeleyde,ves}
\end{icmlauthorlist}

\icmlaffiliation{berkeleycs}{Department of Computer Science, University of California, Berkeley}
\icmlaffiliation{berkeleyde}{Department of Development Engineering, University of California, Berkeley}
\icmlaffiliation{ves}{VESSL AI}

\icmlkeywords{RNA language models, sparse autoencoders, interpretability}

\vskip 0.3in
]

\makeatletter\gdef\icmlcorrespondingauthor@text{Taehan Kim}\makeatother
\printAffiliationsAndNotice{\icmlEqualContribution}

\begin{abstract}
Deep learning, particularly with the advancement of Large Language Models, has transformed biomolecular modeling, with protein language models such as ESM inspiring emerging RNA language models such as RiNALMo. Recent work has begun applying sparse autoencoders (SAEs) to protein language model representations, exploring representation-level interpretability in biomolecular models. Here, we explore whether SAEs can provide interpretable feature decompositions of RNA language model representations, while also examining their limitations in this setting. We present SAE-RNA, interpretability model that analyzes RiNALMo representations and maps them to known human-level biological features. Rather than claiming definitive biological concept discovery, our study frames SAE-based analysis as a representation-level probe for characterizing how RNA language models organize biological information internally. More broadly, SAE-RNA provides a feature-level framework for comparing RNA groups and identifying sparse representation components associated with RNA family identity or structural context.
\end{abstract}

\section{Motivation}
The application of large language models (LLMs) to biology has accelerated in recent years. For RNA, early efforts focused on task-specific models for secondary structure prediction or function classification. For example, early computational approaches focused on family classification using handcrafted or structural features, 
such as nRC~\citep{fiannaca2017nrc}, which combined structural descriptors with machine learning. ncRDense~\citep{chantsalnyam2021ncrdense} also employed convolutional networks to classify ncRNA families directly from sequence. Deep learning application improved upon this with more transferability across problems. More recently, general-purpose RNA LMs such as RiNALMo have shown that pretrained embeddings from a single model can capture diverse RNA properties and support multiple downstream tasks, including secondary structure prediction, ncRNA classification, and splice-site prediction.

At the same time, interpretability has become a central challenge in biomolecular machine learning. Techniques such as SHAP ~\citep{LundbergLee2017SHAP} and Integrated Gradients ~\citep{Sundararajan2017IG} map predictions back to molecular inputs, for example highlighting important atoms or nucleotides for classification decisions. However, these attribution methods primarily focus on explaining outputs and do not reveal what the model encodes in its internal representations. This is especially challenging for large models.

Understanding hidden representations is crucial. Dissecting how RNA LMs organize biological concepts in their embeddings could improve trustworthiness, align model behavior with known biology, and potentially reveal novel patterns not previously recognized. Moreover, representation-level interpretability offers a path to steering model behavior without costly retraining.

Inspired by recent interpretability efforts in Large Language Models, such as Anthropic’s Neuronpedia \cite{neuronpedia}, which maps concepts to individual neurons, we propose a Sparse Autoencoder (SAE)-based model for RNA. Our method, SAE-RNA, identifies interpretable features within hidden states and links them to biological structures and ncRNA families, creating a bridge between deep representations and human-level biological knowledge.

Given an RNA sequence, we extract multiple hidden states across the RNA Language Model and train a Sparse Autoencoder (SAE) on the token space to learn an overcomplete, sparse dictionary of concept units. For each sequence, the SAE yields position-resolved activations that localize concepts along the RNA, while aggregation across tokens provides sequence-level profiles. We then test whether specific sparse concepts align with ncRNA families (e.g., tRNA, riboswitches, snoRNAs) and with structure-aware regions (stems, loops, junctions) or motifs with known functional roles. Our model reveals that RNA language model embeddings are organized into interpretable, reusable concepts that (i) recur within RNA families and (ii) concentrate in structurally meaningful regions.

\section{Related Works}

\subsection{Interpretability in Large Language Models}
Interpretability has been extensively studied in the context of natural language models.
Sparse autoencoders (SAEs) map dense hidden states into higher-dimensional sparse features, revealing disentangled and interpretable concepts~\citep{huben2024sparse}. OpenAI’s neuron-explainer~\citep{OpenAINEURON2023} and Anthropic’s analysis of feature steering~\citep{AnthropicMind2024}
further demonstrate that individual neurons or sparse features can encode semantic concepts and that targeted interventions can systematically steer model behavior.
These efforts highlight the promise of representation-level techniques for both understanding and controlling model internals.

\subsection{Interpretability in Protein Language Models}
Recent work extends these ideas to biomolecular modeling.
InterPLM investigates how protein language model embeddings align with biological categories such as domains and families, and studies concept-level attributions~\citep{InterPLM2025}.
In parallel, sparse autoencoders trained on protein LM representations uncover features corresponding to secondary-structure elements and functional motifs~\citep{ProteinSAE2025},
supporting the utility of sparse feature discovery in scientific domains.

\subsection{RNA Language Models}
RNA language models (e.g., RiNALMo) show promise for downstream tasks, yet it is unclear how their embeddings internally organize biological features.
Our work addresses this gap by adapting SAE-based interpretability to RNA embeddings, probing whether sparse features align with ncRNA families, structure-aware regions, and conserved functional motifs.
This complements input-attribution methods such as SHAP~\citep{LundbergLee2017SHAP} and Integrated Gradients~\citep{Sundararajan2017IG}, which primarily explain outputs rather than hidden representations.
We position our approach alongside advances in large-scale biomolecular LMs (e.g., ESM~\citep{rives2021ESM}) and practical cheminformatics pipelines that visualize attributions on molecular structure~\citep{MolPipelineBASF}.

\section{Methods}

\subsection{Overview}

\begin{figure}[h]
\vspace{.1in}
\centering
\includegraphics[width=\columnwidth]{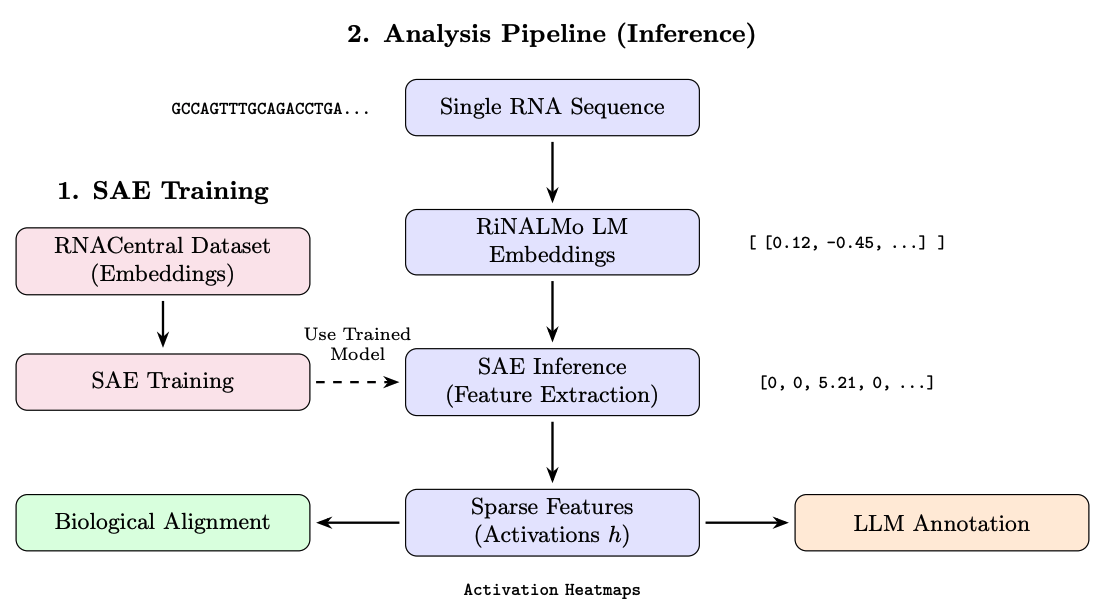}
\vspace{.3in}
\caption{The Analysis Overview. (1) A sparse autoencoder (SAE) is trained offline on embeddings from the RNACentral dataset with balanced family groups. (2) The trained SAE is then used in an analysis pipeline to extract interpretable features for each single RNA sequences.}
\end{figure}

We design SAE-RNA, an interpretability model that probes RNA language model embeddings through a Sparse Autoencoders (SAEs). 
The model proceeds in three stages: (i) extraction of hidden states from a pretrained RNA LM,  RiNALMo, (ii) training of overcomplete SAEs on token-level embeddings from selected layers, and (iii) mapping of discovered sparse features to biological categories and annotations including ncRNA families, secondary-structure regions, and conserved motifs. This design allows us to examine SAE activations at both the motif level and the family level.

\subsection{Embedding Extraction}
We employ RiNALMo~\citep{penic2025rinalmo}, a 650M-parameter RNA language model trained on RNACentral dataset. 
We selected RiNALMo because it is the largest publicly available RNA LM and has been shown to generalize across both RNA families and structural properties. 
For feature extraction, we use sequences from RNACentral,\footnote{\url{https://huggingface.co/datasets/multimolecule/rnacentral}} 
which ensures alignment with RiNALMo’s training distribution and reduces out-of-distribution effects. 
This pairing of model and dataset provides both high-capacity embeddings and training-data consistency, making it well-suited for downstream interpretability analysis. 
We use the HuggingFace implementation by Multimolecule.\footnote{\url{https://huggingface.co/multimolecule/rinalmo-giga}}  

For each RNA sequence, we extract hidden states from multiple transformer layers (\texttt{[1, 9, 18, 24, 30, 33]}). 
Each sequence is represented as a token-level embedding matrix $(L, d)$, where $L$ is the sequence length and $d=1280$ is the embedding dimension. 
These embeddings are standardized and serve as the training input for the SAEs.

\subsection{Sparse Autoencoder Training}
Following prior interpretability works in language models, 
we train a traditional overcomplete SAEs to decompose dense embeddings $x \in \mathbb{R}^d$ into sparse features $f \in \mathbb{R}^k$, where $k \gg d$. 
Our SAE consists of a linear encoder, ReLU activation, and linear decoder:
\[
f = \mathrm{ReLU}(W_e x + b), \quad \hat{x} = W_d f + c.
\]
Unlike tied-weight autoencoders, we use untied encoder and decoder weights.  
Weights are initialized with Kaiming (encoder) and Xavier (decoder) initialization.  

The objective combines reconstruction and sparsity:
\[
\mathcal{L} = \|x - \hat{x}\|_2^2 + \lambda \|f\|_1,
\]
with $\lambda$ controlling feature sparsity. In practice we set $\lambda = 3 \times 10^{-3}$, $lr = 1 \times 10^{-3}$, and $weight decay = 1 \times 10^{-4}$.

\subsection{Training Procedure}
We train one SAE per RiNALMo layer. Each layer’s token activations are batched with size 1024. 
Training uses AdamW optimizer (learning rate $10^{-3}$, weight decay $10^{-4}$), cosine annealing scheduling, and gradient clipping at a norm 1.0. 
Each model is trained for 10 epochs.  
We monitor mean squared reconstruction error, average L1 penalty, and effective sparsity (mean number of active features per token).  
Trained SAEs and dataset standardization statistics are checkpointed for downstream analysis. Layer-wise reconstruction-loss convergence curves are provided in Appendix~\ref{app:training-curves} (Figure~\ref{fig:training-curves-appendix}).
\subsection{Feature Localization}
For each sequence, the trained SAE produces position-resolved activations $h_{i,j}$, where $h_{i,j}$ denotes the activation of feature $j$ at token $i$.  
Aggregating activations across tokens yields sequence-level profiles, enabling the localization of sparse concepts at both the nucleotide level (e.g., stems vs.\ loops) and the family level (e.g., tRNAs vs.\ riboswitches).  
We focus on features that exhibit consistent activation patterns across subsets of sequences.
\subsection{Biological Alignment and Evaluation using \href{https://huggingface.co/datasets/multimolecule/bprna-90}{bpRNA-90} and RNAcentral}

We evaluate the interpretability of discovered features along two complementary axes:

\paragraph{(1) Structural and motif alignment (bpRNA-90).}  
The first analysis focuses on mapping SAE features to structural and motif-level biological elements. We use the 2000 samples from sequence length cut off of 2000 from the \texttt{bnRNA-90} dataset  \footnote{\url{https://huggingface.co/datasets/multimolecule/bprna-90}}, which provides nucleotide sequences, secondary structures, as well as structural and functional annotations.  Unlike protein models, where amino acid sequences can often be mapped directly to motifs, RNA features are less reliably inferred from sequence alone. It requires both sequence and structural context for comprehensive interpretation. Thus, we incorporate the precise structural regions where activations occur.  
Structural annotations in bpRNA-90 categorize regions as:  
E (External loop), S (Stem), H (Hairpin loop), I (Internal loop), M (Multi-loop), B (Bulge), X (Ambiguous/undetermined), and K (Pseudoknot).  

This enables us to map highly activated spans not only to their nucleotide sequences but also to their structural and functional contexts. For example, we can extract out the sample information in the structure below for labeling.

\begin{Verbatim}[breaklines=true,breakanywhere=true,fontsize=\footnotesize]
bpRNA_sample_format | len=n | 
spans=[(50, 51), (144, 146)...] | 
nts=['GG', 'CGG', ...] | 
struct=['MM', 'SSS', ...] | 
func=['KK', 'NNN', ...]
\end{Verbatim}

\paragraph{(2) Family-level alignment (RNAcentral).}  
The second analysis shifts focus to noncoding RNA families. Using a per-family balanced sample of $\sim$3,000 sequences from RNAcentral, we test whether specific SAE features preferentially activate in distinct ncRNA families such as tRNAs, riboswitches, or snoRNAs. This enables us to track layer-wise learning trends at the family level, providing a complementary perspective to the structural motif analysis above.

\paragraph{Summary.}  
Together, these analyses allow us to characterize features at two scales:  
(i) motif- and structure-resolved features using bpRNA-90, where activations are mapped to precise structural and biological labels, and  
(ii) family-level activation trends using RNAcentral, where we assess whether features capture higher-level functional organization across ncRNA classes. 

\subsection{Feature Annotation via Prompting}
To systematically convert sparse autoencoder features into interpretable labels, we employ large language model (GPT-5) prompting with structured templates. 
Each prompt provides the following information:

\begin{itemize}
    \item \textbf{Activation statistics:} number of sequences with activations, family distribution, base composition, $n$-mer counts, positional bias, and island counts.  
    \item \textbf{Example spans:} contiguous subsequences where activations are strongest, extracted via percentile thresholds. Because RNA motifs cannot be reliably inferred from sequence alone (unlike many protein motifs), spans include both \textit{nucleotide sequences} and \textit{bpRNA structural annotations} (E=External, S=Stem, H=Hairpin, I=Internal, M=Multiloop, B=Bulge, X=Ambiguous, K=Pseudoknot).  
    \item \textbf{Motif reference list:} a curated catalog of canonical RNA motifs (e.g., GNRA, UNCG, kink-turn, Shine–Dalgarno, sarcin–ricin).  
\end{itemize}

\noindent An illustrative prompt snippet from the provided list is shown below:

\begin{Verbatim}[breaklines=true,breakanywhere=true,fontsize=\footnotesize]
Feature ID: 6510
Aggregate stats:
- Sequences with activations: 51
- Base composition: A=0.10, C=0.15,..
- Top 2-mers: GG, CG, GC
- Top 3-mers: GGG, CGG, GGC
- Positional bias: mean=0.24 (towards 5')

Example spans list (truncated):
- bpRNA_sample_1 | len=366 | 
  spans=[(50, 51), (144, 146), (156, 158)] |
  nts=['NT1', 'NT2', 'NT3'] |
  struct=['ST1', 'ST2', 'ST3'] |
  func=['F1', 'F2', 'F3']
- bpRNA_sample_2 | len=315 |
  spans=[(112, 114), (136, 140)] |
  nts=['NT4', 'NT5'] |
  struct=['ST4', 'ST5'] |
  func=['F4', 'F5']
\end{Verbatim}

The model is instructed to:
\begin{enumerate}
    \item generate a structured multi-bullet description of activation patterns (covering sequence bias, motif recurrence, and dominant structural enrichment), and 
    \item assign a concise shorthand label with rationale that combines sequence and structure (e.g., “AU-rich [S] — Stem-associated AU tracts”).
\end{enumerate}

We standardize the prompt format across layers to ensure reproducibility.  
Features that align with known motifs serve as internal validation, while novel but consistent activation signatures are noted for follow-up.  

Finally, for primary motif-related features (e.g., hairpins, stems), we further validate LLM-generated annotations by cross-checking them against structural mappings presented in the following sections. While exhaustive manual inspection is infeasible, this two-step pipeline—automatic span-based prompting followed by selective human review, provides increased confidence that the discovered features reflect biologically meaningful motifs rather than artifacts of the embedding space. All reported features are drawn from a filtered set \textbf{restricted to those firing on at least 10 distinct sequences}, ensuring that analyses are based on consistently supported activations rather than rare or spurious events.

\subsection{Implementation Details}
All models are implemented in PyTorch. Training and evaluation are conducted on NVIDIA A100 GPUs. 
We fix $d=1280$ (embedding dimension) and $h=10240$ (dictionary size). 
Hyperparameters are selected to balance reconstruction fidelity with interpretability. 
Epoch-level metrics include reconstruction error, mean absolute activation, and sparsity rate.

\section{Results}
\begin{figure}[h]
\vspace{.1in}
\centering
\includegraphics[width=\columnwidth]{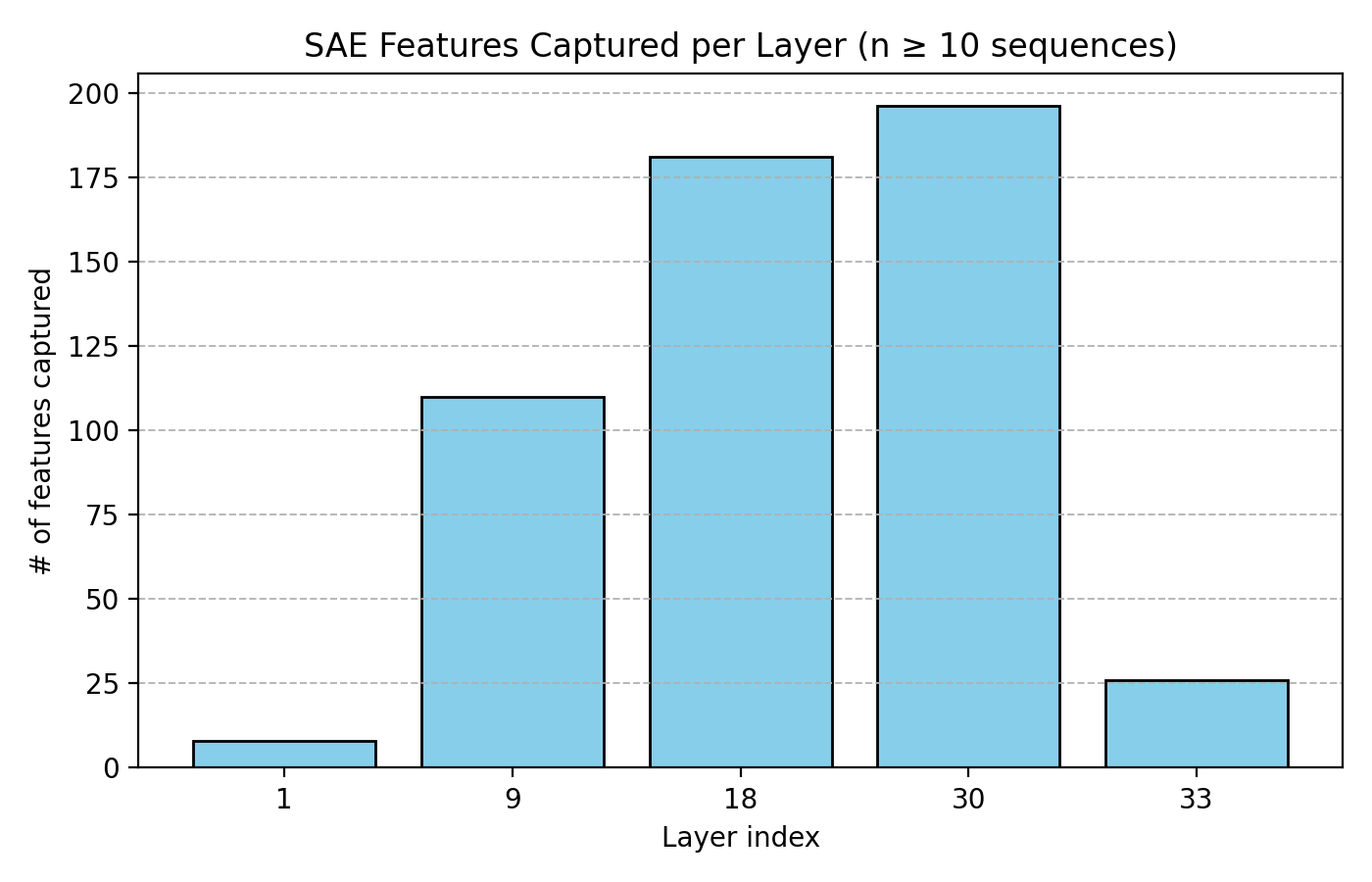}
\vspace{.1in}
\caption{Number of SAE features per layer that were retained after filtering for activations in at least 10 distinct sequences.}
\end{figure}

\subsection{Layer-wise Motif Labeling of Activated Features}

\begin{table}[ht]
\caption{Layer 9: Activated Feature Labels (subset).} \label{layer9-table}
\centering

\begin{tabularx}{\linewidth}{@{}
  >{\raggedright\arraybackslash}p{0.9cm}
  >{\raggedright\arraybackslash}p{2.2cm}
  X
@{}} 
\textbf{ID} & \textbf{LABEL} & \textbf{RATIONALE} \\
\cmidrule(l{4pt}r{4pt}){1-1}\cmidrule(l{4pt}r{4pt}){2-2}\cmidrule(l{4pt}r{4pt}){3-3}
2053 & \makecell[l]{Poly-G [S]\\Stem helix} &
\makecell[l]{Explicit G-runs observed.\\Dominated by S (\(\approx 55\%\)).} \\
2200 & \makecell[l]{Poly-A [H]\\Hairpin loop} &
\makecell[l]{Explicit A-runs observed.\\Dominated by H (\(\approx 54\%\)).} \\
820  & \makecell[l]{Poly-G [E]\\External} &
\makecell[l]{Explicit G-runs observed.\\Dominated by E (\(\approx 89\%\)).} \\
1146 & \makecell[l]{A-rich [H]\\Hairpin loop} &
\makecell[l]{Strong A-rich spans.\\Dominated by H (\(\approx 83\%\)).} \\
8004 & \makecell[l]{Poly-A\\(no dominant \\ stucture)} &
\makecell[l]{A-runs present, but \\ structural classes mixed;\\ max H (\(\approx 40\%\)).} \\
2892 & \makecell[l]{GC-rich [S]\\Stem helix} &
\makecell[l]{GC-rich pattern, strong\\enrichment in stems (\(\approx 75\%\)).} \\
9374 & \makecell[l]{U-rich [S]\\Stem helix} &
\makecell[l]{U-rich spans observed.\\Dominated by S (\(\approx 54\%\)).} \\
\end{tabularx}
\end{table}

\begin{table}[ht]
\caption{Layer 18: Activated Feature Labels (subset)} \label{layer18-table}
\centering

\begin{tabularx}{\linewidth}{@{}
  >{\raggedright\arraybackslash}p{0.9cm}
  >{\raggedright\arraybackslash}p{2.2cm}
  X
@{}} 
\textbf{ID} & \textbf{LABEL} & \textbf{RATIONALE} \\
\cmidrule(l{4pt}r{4pt}){1-1}\cmidrule(l{4pt}r{4pt}){2-2}\cmidrule(l{4pt}r{4pt}){3-3}
7783  & \makecell[l]{Poly-C [H]\\Hairpin loop} &
\makecell[l]{Explicit C-runs observed.\\Dominated by H (\(\approx 80\%\)).} \\
4459  & \makecell[l]{Poly-G [S]\\Stem helix} &
\makecell[l]{Explicit G-runs observed.\\Dominated by S (\(\approx 65\%\)).} \\
4618  & \makecell[l]{Poly-G [S]\\Stem helix} &
\makecell[l]{Explicit G-runs observed.\\Dominated by S (\(\approx 72\%\)).} \\
219   & \makecell[l]{Poly-U [S]\\Stem helix} &
\makecell[l]{Explicit U-runs observed.\\Dominated by S (\(\approx 50\%\)).} \\
726   & \makecell[l]{Poly-A [S]\\Stem helix} &
\makecell[l]{Explicit A-runs observed.\\Dominated by S (\(\approx 67\%\)).} \\
10217 & \makecell[l]{U-rich [H]\\Hairpin loop} &
\makecell[l]{U-rich spans observed.\\Dominated by H (\(\approx 100\%\)).} \\
6417  & \makecell[l]{Poly-C [H]\\Hairpin loop} &
\makecell[l]{Explicit C-runs observed.\\Dominated by H (\(\approx 59\%\)).} \\
10216 & \makecell[l]{A-rich\\(no dominant \\ stucture)} &
\makecell[l]{A-rich spans; mixed structures,\\max S (\(\approx 43\%\)).} \\
\end{tabularx}
\end{table}

\begin{figure}[t]
\centering
\includegraphics[width=\columnwidth]{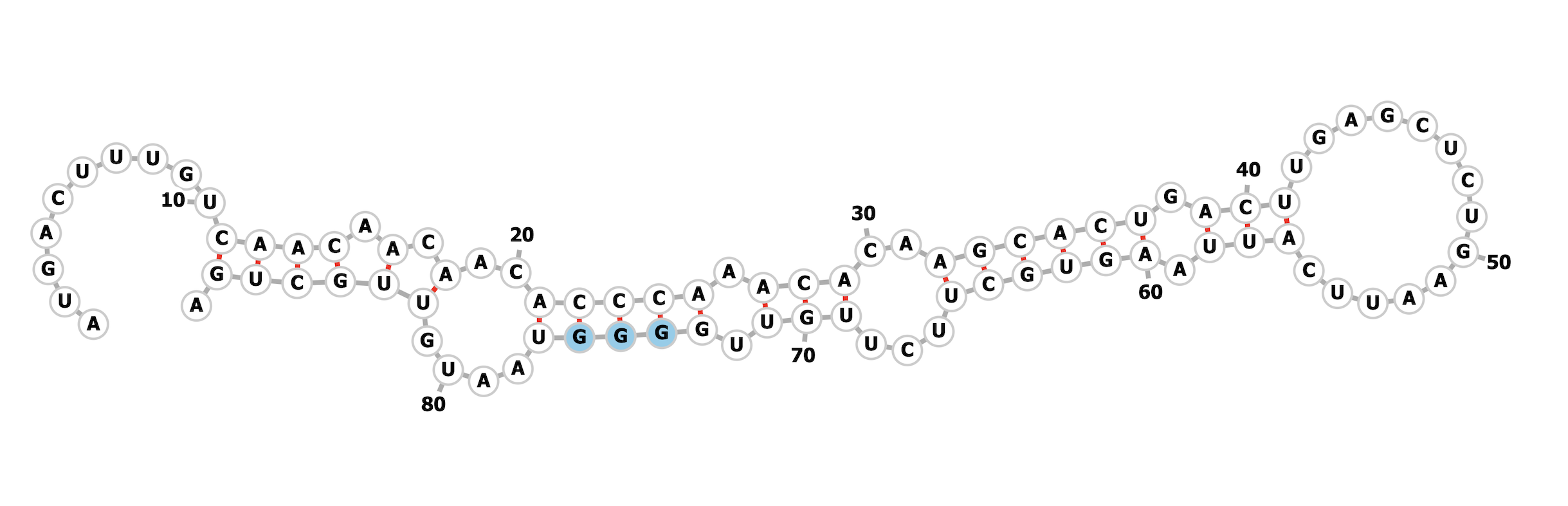}

\includegraphics[width=\columnwidth]{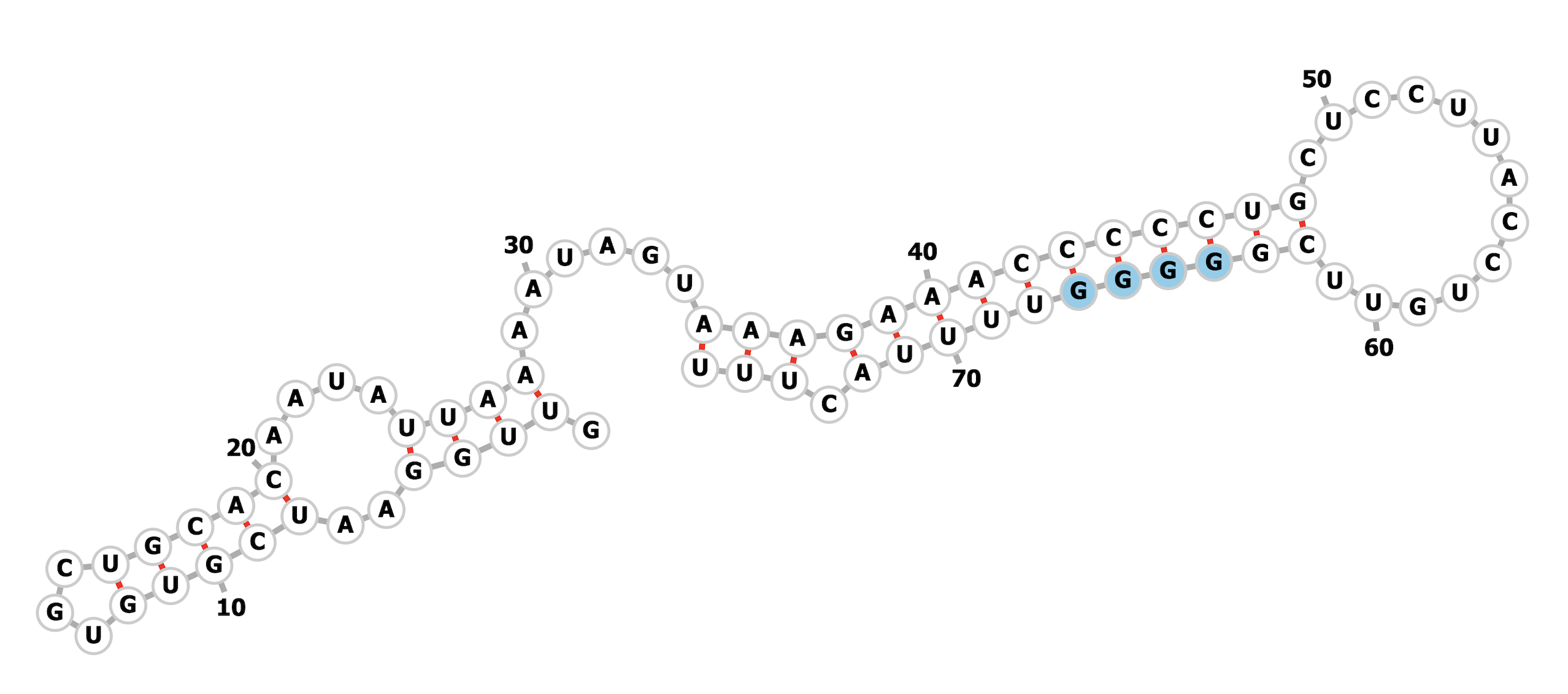}

\caption{Activated sequence of bpRNA-RFAM-25894 and bpRNA-RFAM-42383  at token level by feature 2053: (Stem).}
\label{fig:activated-sequence-stem}
\end{figure}

\begin{figure}[h]
\centering
\includegraphics[width=0.7\columnwidth]{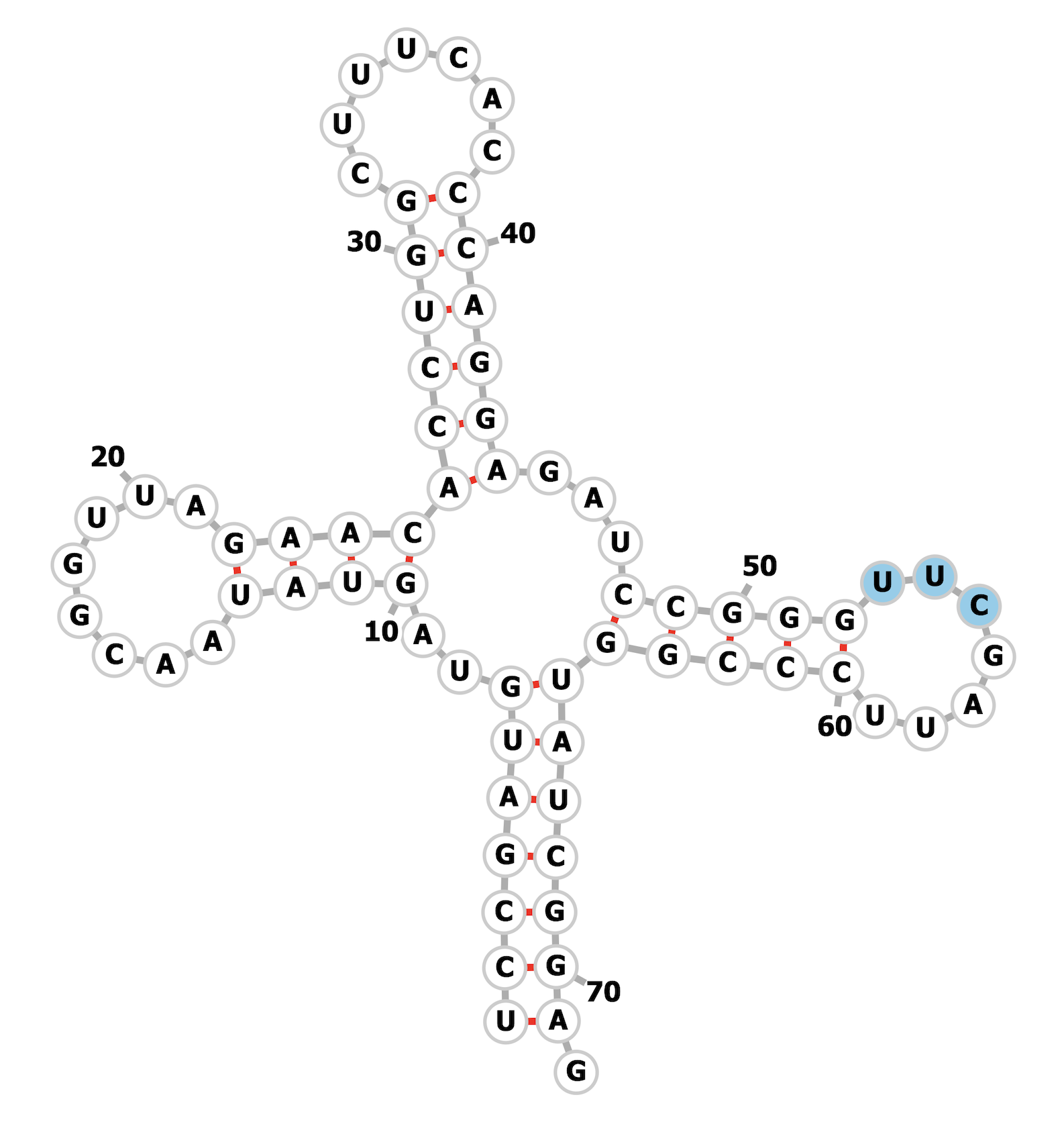}
\caption{Activated sequence of bpRNA-CRW-29143 in token-level by feature 7783: (Hairpin)}
\label{fig:activated-sequence-hairpin}
\end{figure}

\textbf{Observation.}
We analyzed the token-level sequences on which each feature is activated. For every feature ID, we annotated a \emph{rationale} and an associated \emph{label}, following the scheme summarized in Table~\ref{layer9-table} and Table~\ref{layer18-table}. We then conducted a deeper investigation by visualizing these rationales to examine their correspondence with RNA secondary-structure contexts. This analysis revealed that many features fire on specific structural elements—most notably stems and hairpins—when the corresponding sequence patterns are present (see Fig.~\ref{fig:activated-sequence-stem} and Fig.~\ref{fig:activated-sequence-hairpin}.)

\subsection{Layer-wise Emergence of RNA Functional Type Selective Features}

\textbf{Setup.}
We analyze layers $\{1, 9, 18, 24, 30, 33\}$. For each RNA type and layer, we select the top-$k$ ($k{=}5$) most-activated channels and visualize the union heatmap after per-feature normalization and max aggregation across positions/samples. 
The main figure contrasts Layer~1 (L1), Layer~18 (L18), and Layer~33 (L33).

\textbf{Observation.}
As we move from L1 to later layers, activations shift from diffuse and widely shared to sparser, higher-contrast patterns with strong peaks on a small subset of channels per RNA type. 
This \emph{is not linear across depth}: L1 exhibits the lowest sparsity, and from L18 onward sparsity and type selectivity are markedly higher and remain elevated (with mild fluctuations) through L33. Visually, this appears as a reduction in background ``noise'' after L1 and a concentration of activation on a few type-preferential features in deeper layers. 

\textbf{Main hypothesis.}
(1) After \emph{L1 $\rightarrow$ jump in effective denoising}: downstream layers suppress broadly distributed, low-informative responses, producing a step-like increase in sparsity after L1.\\
(2) \emph{Explicit sparsity thereafter}: repeated nonlinearity and normalization yield higher peak responses so that only a few channels remain prominent for each RNA type from L18 onward.\\
(3) \emph{Visualization}: max aggregation amplifies strong in-type peaks and deemphasizes weaker off-type responses, making the late-layer selectivity visually salient.

\begin{figure}[H]
\centering
\includegraphics[width=0.95\linewidth]{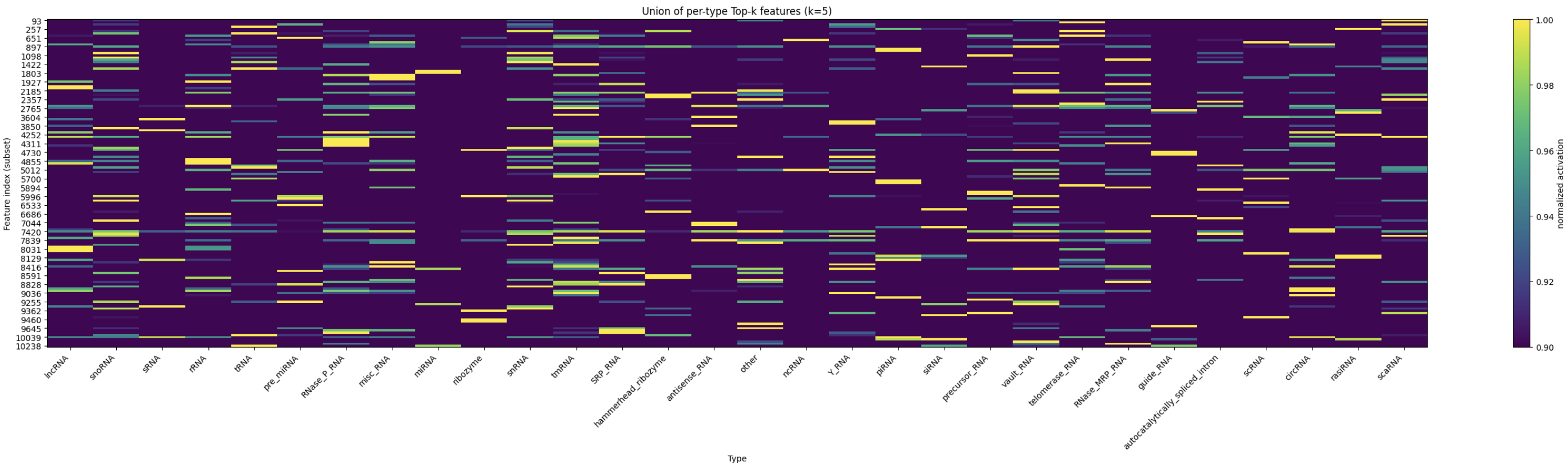}\\[-2mm]
\includegraphics[width=0.95\linewidth]{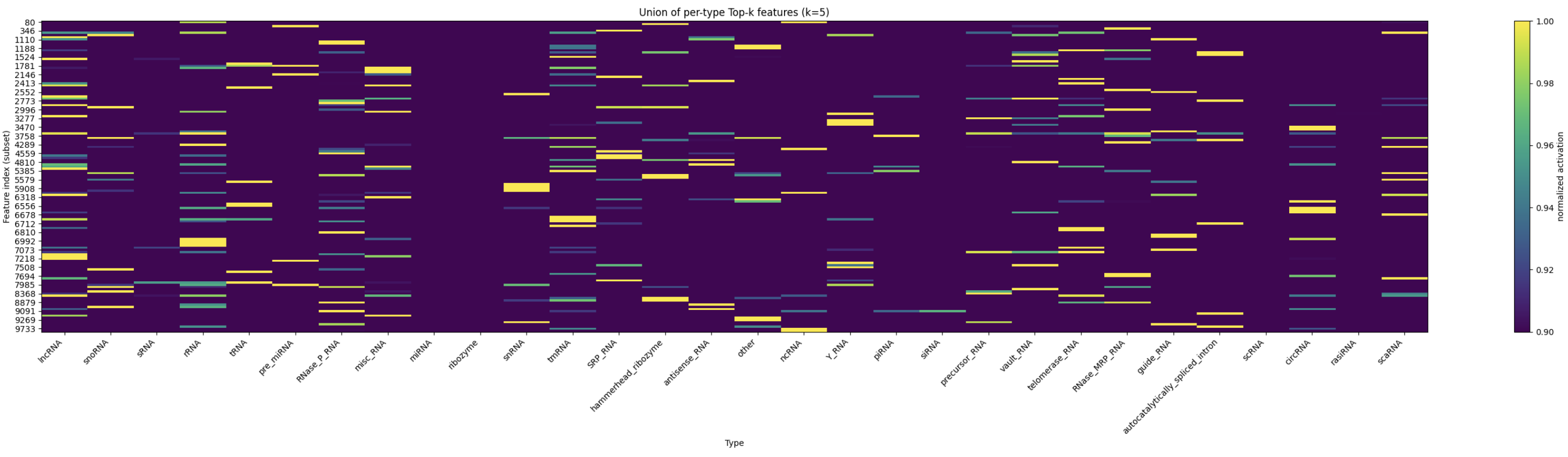}\\[-2mm]
\includegraphics[width=0.95\linewidth]{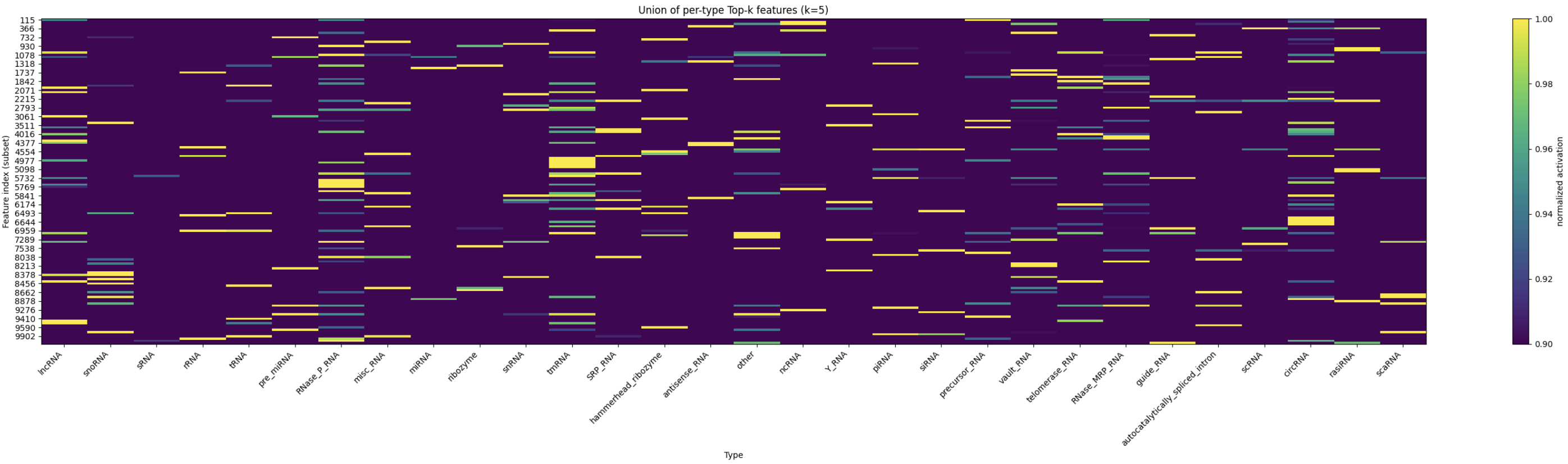}
\caption{\textbf{Union of per-type top-$k$ features} for L1 (top), L18 (middle), and L33 (bottom). 
Color shows normalized activation; the $y$-axis indexes selected feature channels and the $x$-axis enumerates RNA types. 
L1 shows the lowest sparsity; from L18 onward, patterns are markedly sparser and more type-selective. }
\label{fig:layer-heatmaps}
\end{figure}

\textbf{Explanation.}
From L1 to deeper layers, we observe a clear strengthening of sparsity and type selectivity. 
In particular, L1 shows the lowest sparsity, while L18 and beyond exhibit markedly sparser, high-contrast patterns with strong peaks concentrated on a small subset of channels per RNA type. 
This progression is directly visible in the heatmaps (Fig.~\ref{fig:layer-heatmaps}; L1, L18, L33), where background activation diminishes after L1 and feature-wise selectivity becomes more pronounced in later layers.

\section{Discussion and Limitations}
In this work, we explored the potential of sparse autoencoders (SAEs) as a representation-level interpretability tool for RNA language models. Our analyses suggest that selected SAE features can exhibit some activation patterns aligned with known RNA sequence and structural annotations, including stem and hairpin associated regions, as well as RNA family activation patterns across layers. These results indicate that SAEs may potentially provide a useful feature-level lens for probing how RNA language models organize biological information internally.

At the same time, our results should be interpreted conservatively. Although SAEs have been proposed as a way to recover sparse and potentially interpretable features from large models, our study does not establish that the learned features correspond to definitive biological concepts. In the current setting, the most reliable interpretation comes from comparing feature activations against "already known" annotations, such as secondary-structure labels and ncRNA family labels. Therefore, SAE-RNA should be viewed as an exploratory analysis framework rather than a validated biological discovery method.

Several methodological limitations remain. First, feature interpretation depends strongly on the availability and quality of known biological annotations. Without external labels or orthogonal validation, it is difficult to determine whether an activated feature reflects a meaningful motif, a correlated sequence feature, or noise in the representation. Second, RNA sequences can be long and heterogeneous, making sparse activation spikes difficult to interpret. In long sequences, isolated high activations may correspond to true localized motifs, but they may also arise from background noise, thresholding artifacts, or length-dependent effects. This makes normalization across sequences of different lengths an important unresolved issue. Third, feature-level conclusions may depend on choices such as the SAE sparsity penalty, activation threshold, top-$k$ feature selection, aggregation method, and layer selection. These choices can affect which features appear salient and how strongly they seem to align with biological annotations. Currently, there are no systematic choices and the choices are not yet standardized.

\textbf{Future Work.}
These limitations suggest several directions for future work. More rigorous validation could include perturbation tests and motif-masking experiments. We also explored preliminary extensions, including evaluation on synthetic motif-perturbation data and preliminary downstream uses of SAE-derived features for mutation-associated motif analysis. However, we found that the current feature activations were not yet sufficiently stable, normalized, or validated to support reliable downstream conclusions. In particular, sparse activation spikes in long and variable-length RNA sequences remain difficult to distinguish from noise.

Future work should therefore evaluate whether SAE-derived features improve predictive performance, sample efficiency, or interpretability beyond standard embedding-based baselines only after these methodological issues are addressed. Overall, our results support the potential of SAE-based analysis for RNA language models, while emphasizing that feature stability, length-normalization procedures that preserve localized motif signals, and signal-versus-noise validation must be resolved before SAE features can be used as reliable biological markers or discovery tools.

\section{Conclusion}

We introduced \textbf{SAE-RNA}, an exploratory framework for applying sparse autoencoders to hidden representations from RNA language models. Rather than treating SAE features as definitive biological concepts, our goal was to examine whether sparse feature decompositions can provide interpretable probes of RNA language model representations. Trained on token-level RiNALMo embeddings across multiple layers, SAE-RNA produces position-aligned sparse activations.

Using \textit{bpRNA-90}, we observed that selected features activate in recognizable sequence and secondary-structure contexts, including stem and hairpin associated regions with clear sequence biases that map onto secondary-structure contexts (Fig.~\ref{fig:activated-sequence-stem}, Fig.~\ref{fig:activated-sequence-hairpin}). Complementary analyses using \textit{RNAcentral} suggest that deeper RiNALMo layers exhibit more concentrated RNA-type-selective activation patterns than early layers (Fig.~\ref{fig:layer-heatmaps}). Together, these results suggest that SAE-based analysis can give sparse representation components associated with known RNA structural and family-level annotations. However, they do not yet establish SAE features as newly discovered biological concepts. 

Important limitations remain. First, our SAE training was constrained by computational resources: we trained on 10k RNA sequences, whereas scaling to millions of sequences would likely provide broader coverage of RNA families, motifs, and structural contexts. Second, the learned features may be sensitive to SAE-specific choices, including the sparsity penalty, activation threshold, layer selection, and aggregation strategy. Third, long and variable-length RNA sequences make it difficult to distinguish true localized motif signals from noisy activation spikes, and normalization across sequences of different lengths remains an unresolved methodological challenge. Future work should therefore scale SAE training to larger RNA corpora, evaluate feature stability across different hyperparameter settings, and validate candidate features through synthetic perturbation analysis, motif masking, and downstream task experiments before suggesting SAE features as reliable biological markers for real downstream tasks.

\section{Acknowledgement}
This work was made possible by the \textbf{VESSL AI} GPU platform, with the support from A100 GPUs and external storage. We thank VESSL AI for their support.


\bibliographystyle{icml2026}
\bibliography{references}

@article{fiannaca2017nrc,
  title={nRC: non-coding RNA Classifier based on structural features},
  author={Fiannaca, Antonino and La Rosa, Massimo and La Paglia, Laura and Rizzo, Riccardo and Urso, Alfonso},
  journal={BioData mining},
  volume={10},
  number={1},
  pages={27},
  year={2017},
  publisher={Springer}
}

@article{chantsalnyam2021ncrdense,
  title={ncRDense: a novel computational approach for classification of non-coding RNA family by deep learning},
  author={Chantsalnyam, Tuvshinbayar and Siraj, Arslan and Tayara, Hilal and Chong, Kil To},
  journal={Genomics},
  volume={113},
  number={5},
  pages={3030--3038},
  year={2021},
  publisher={Elsevier}
}

@misc{neuronpedia,
    title = {Neuronpedia: Interactive Reference and Tooling for Analyzing Neural Networks},
    year = {2023},
    note = {Software available from neuronpedia.org},
    url = {https://www.neuronpedia.org},
    author = {Lin, Johnny}
}

@inproceedings{huben2024sparse,
  title={Sparse autoencoders find highly interpretable features in language models},
  author={Huben, Robert and Cunningham, Hoagy and Smith, Logan and Ewart, Aidan and Sharkey, Lee},
  booktitle={International Conference on Learning Representations},
  volume={2024},
  pages={7827--7845},
  year={2024}
}

@techreport{OpenAINEURON2023,
  title        = {Language Models Can Explain Neurons in Language Models},
  author       = {Bills, Steven and Cammarata, Nick and Mossing, Dan and Tillman, Henk and Gao, Leo and Goh, Gabriel and Sutskever, Ilya and Leike, Jan and Wu, Jeff and Saunders, William},
  year         = {2023},
  institution = {OpenAI},
  howpublished = {\url{https://openaipublic.blob.core.windows.net/neuron-explainer/paper/index.html}},
  note         = {OpenAI, published May 9, 2023}
}

@article{AnthropicMind2024,
   title={Scaling Monosemanticity: Extracting Interpretable Features from Claude 3 Sonnet},
   author={Templeton, Adly and Conerly, Tom and Marcus, Jonathan and Lindsey, Jack and Bricken, Trenton and Chen, Brian and Pearce, Adam and Citro, Craig and Ameisen, Emmanuel and Jones, Andy and Cunningham, Hoagy and Turner, Nicholas L and McDougall, Callum and MacDiarmid, Monte and Freeman, C. Daniel and Sumers, Theodore R. and Rees, Edward and Batson, Joshua and Jermyn, Adam and Carter, Shan and Olah, Chris and Henighan, Tom},
   year={2024},
   journal={Transformer Circuits Thread},
   url={https://transformer-circuits.pub/2024/scaling-monosemanticity/index.html}
    }

@article{InterPLM2025,
  title={InterPLM: discovering interpretable features in protein language models via sparse autoencoders},
  author={Simon, Elana and Zou, James},
  journal={Nature methods},
  volume={22},
  number={10},
  pages={2107--2117},
  year={2025},
  publisher={Nature Publishing Group US New York}
}

@article{ProteinSAE2025,
  title={Sparse autoencoders uncover biologically interpretable features in protein language model representations},
  author={Gujral, Onkar and Bafna, Mihir and Alm, Eric and Berger, Bonnie},
  journal={Proceedings of the National Academy of Sciences},
  volume={122},
  number={34},
  pages={e2506316122},
  year={2025},
  publisher={National Academy of Sciences}
}

@article{LundbergLee2017SHAP,
  title={A unified approach to interpreting model predictions},
  author={Lundberg, Scott M and Lee, Su-In},
  journal={Advances in neural information processing systems},
  volume={30},
  year={2017}
}

@inproceedings{Sundararajan2017IG,
  title={Axiomatic attribution for deep networks},
  author={Sundararajan, Mukund and Taly, Ankur and Yan, Qiqi},
  booktitle={International conference on machine learning},
  pages={3319--3328},
  year={2017},
  organization={PMLR}
}

@article{MolPipelineBASF,
  title={MolPipeline: a python package for processing molecules with RDKit in scikit-learn},
  author={Sieg, Jochen and Feldmann, Christian W and Hemmerich, Jennifer and Stork, Conrad and Sandfort, Frederik and Eiden, Philipp and Mathea, Miriam},
  journal={Journal of Chemical Information and Modeling},
  volume={64},
  number={24},
  pages={9027--9033},
  year={2024},
  publisher={ACS Publications}
}

@article{rives2021ESM,
  title={Biological structure and function emerge from scaling unsupervised learning to 250 million protein sequences},
  author={Rives, Alexander and Meier, Joshua and Sercu, Tom and Goyal, Siddharth and Lin, Zeming and Liu, Jason and Guo, Demi and Ott, Myle and Zitnick, C Lawrence and Ma, Jerry and others},
  journal={Proceedings of the national academy of sciences},
  volume={118},
  number={15},
  pages={e2016239118},
  year={2021},
  publisher={National Academy of Sciences}
}

@article{penic2025rinalmo,
  title={Rinalmo: General-purpose rna language models can generalize well on structure prediction tasks},
  author={Peni{\'c}, Rafael Josip and Vla{\v{s}}i{\'c}, Tin and Huber, Roland G and Wan, Yue and {\v{S}}iki{\'c}, Mile},
  journal={Nature Communications},
  volume={16},
  number={1},
  pages={5671},
  year={2025},
  publisher={Nature Publishing Group UK London}
}

\clearpage
\appendix
\thispagestyle{empty}

\onecolumn
\section*{Supplementary Material}
\section{LLM BASED ANNOTATION}
The following prompts were provided to the LLM to guide its annotation process consistently. The model was also provided with a consistent and reliable reference for motif identification.
\subsection{MOTIF REFERENCE}

\begin{mdframed}[
  linewidth=0.8pt,     
  linecolor=black,     
  roundcorner=6pt,     
  backgroundcolor=white, 
  innertopmargin=6pt,  
  innerbottommargin=6pt]
\begin{Verbatim}[breaklines=true,breakanywhere=true,fontsize=\footnotesize]
MOTIF_REFERENCE = """
Here is a non-exhaustive list of common RNA sequence/structural motifs you may refer to 
when labeling features. If the activations match one of these, use its canonical name. 
If not, describe the best-fitting pattern (e.g., "poly-A tracts", "AU-rich runs"). 

This list is for reference only — you are not restricted to it.

Motif ID — Description
ANYA — ANYA tetraloop
AUF1_binding - AUF1 binding site
C-loop — C-loop motif
CRC_binding - CRC binding motif
CsrA_binding - CsrA / RsmA binding motif
CUYG - CUYG tetraloop
Domain-V - Splicing domain V
GNRA - GNRA tetraloop
HuR_binding - HuR binding site
k-turn-1 - Kink turn 1 (3′ bulge)
k-turn-2 — Kink turn 2 (5′ bulge)
pK-turn - RNase P RNA pseudoknot kink turn
RBS_B_subtilis — Shine-Dalgarno (B. subtilis subtype)
RBS_E_coli — Shine-Dalgarno (E. coli subtype)
RBS_H_pylori - Shine-Dalgarno (H. pylori subtype)
right_angle-2 — Right angle motif 2
right_angle-3 — Right angle motif 3
sarcin-ricin-1 — Sarcin-ricin motif 1
sarcin-ricin-2 — Sarcin-ricin motif 2
SRP_S_domain — Signal recognition particle small S-domain
tandem-GA — Tandem GA / AG loop
Terminator1 — Rho independent terminator 1
Terminator2 — Rho independent terminator 2
T-loop — T-loop motif
TRIT — Tuberculosis rho independent terminator
twist_up — Twist up motif
UAA_GAN — UAA / GAN internal loop motif
UMAC — UMAC tetraloop
UNCG — UNCG tetraloop
U-turn — U-turn motif
vapC_target — vapC ribonuclease target
docking_elbow — Docking elbow of tRNA/RNase P/T-box leaders
VTS1_binding — VTS1 binding site
Roquin_binding — Roquin binding site"""
\end{Verbatim}
\end{mdframed}

\subsection{FEATURE ANNOTATION}
\begin{mdframed}[
  linewidth=0.8pt,     
  linecolor=black,     
  roundcorner=6pt,     
  backgroundcolor=white, 
  innertopmargin=6pt,  
  innerbottommargin=6pt]
\begin{Verbatim}[breaklines=true,breakanywhere=true,fontsize=\footnotesize]
FEATURE_PROMPT_TEMPLATE = f"""You are labeling what a neural feature 
(from a sparse autoencoder trained on RNA LM states) is detecting.

Use ONLY the information below. Do not invent RNA structure unless implied by motifs.

{MOTIF_REFERENCE}
TASK A — Pattern description (3–5 bullets):
Start with: "The activation patterns are characterized by:"
• Sequence composition (AU-rich, GC-rich, poly-A, GU-rich).  
• Where in the sequence activations occur (islands, dispersed, positional bias 5′→3′).  
• Recurring motifs (2–3-mers, longer runs).  
• If evidence suggests known motifs from the reference list, mention them explicitly.  
• Differences across RNA families (e.g., stronger in ncRNAs).  
• If there is a recurring novel motif the model is activating on, 
note it explicitly and describe it crisply.


TASK B — Feature label + rationale:
- **Label**: Concise motif name if recognized, or descriptive shorthand otherwise.  
- **Rationale**: In 2–4 sentences, justify the label based on spans, 
k-mers, and activation profile.  
Mention both the dominant base composition and the dominant structural context. 

                    
 If ambiguity is high, state so.
 Decision rule**:
 A feature is only considered "notable" if it shows 
 (a) a clear sequence pattern *and* 
 (b) a clear structural context bias. 
 Always report both together in the label and rationale.


Feature ID: {{feature_id}}

Aggregate stats across sequences:
- Sequences with activations: {{n_sequences}}
- Family counts: {{family_counts}}
- Base composition inside spans (A/C/G/U): {{inside_base_fractions}}
- Top 2-mers: {{top2mers}}
- Top 3-mers: {{top3mers}}
- Total activation islands: {{total_islands}} (median span length: {{median_span_len}})
- Relative position mean±std (0=5′,1=3′): {{pos_mean}} ± {{pos_std}}

Example spans (truncated to {{n_examples}} sequences):
{{examples_block}}
"""
\end{Verbatim}
\end{mdframed}

\section{HEATMAP OF FEATURES PER LAYER}

\begin{figure}[H]
\centering
\includegraphics[width=0.7\linewidth]{heatmap_per_type_layer1.png}\\[-2mm]
\includegraphics[width=0.7\linewidth]{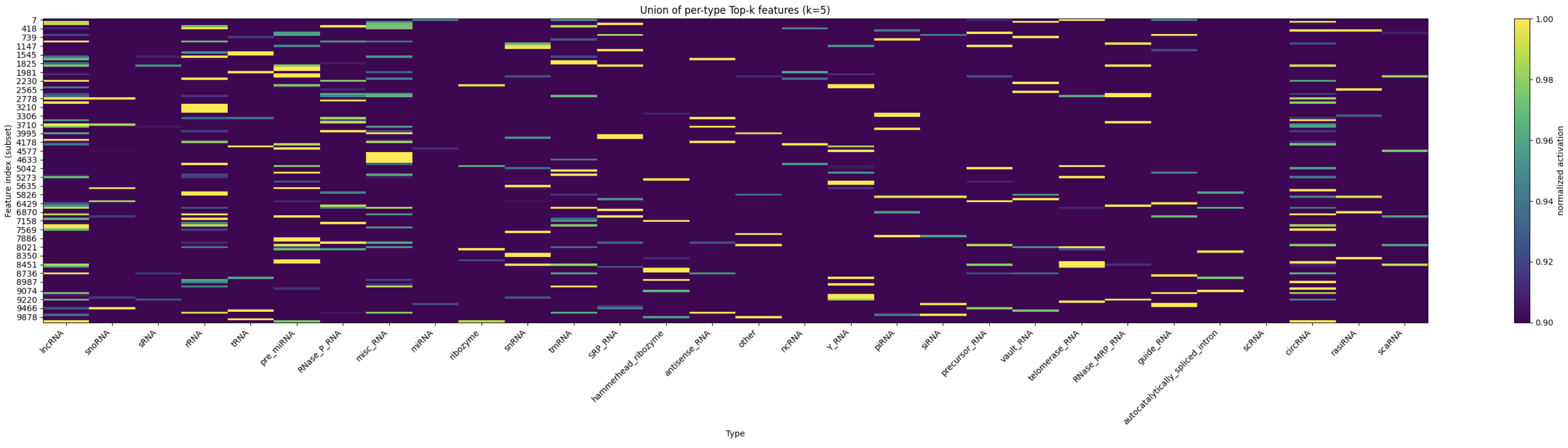}\\[-2mm]
\includegraphics[width=0.7\linewidth]{heatmap_per_type_layer18.png}\\[-2mm]
\includegraphics[width=0.7\linewidth]{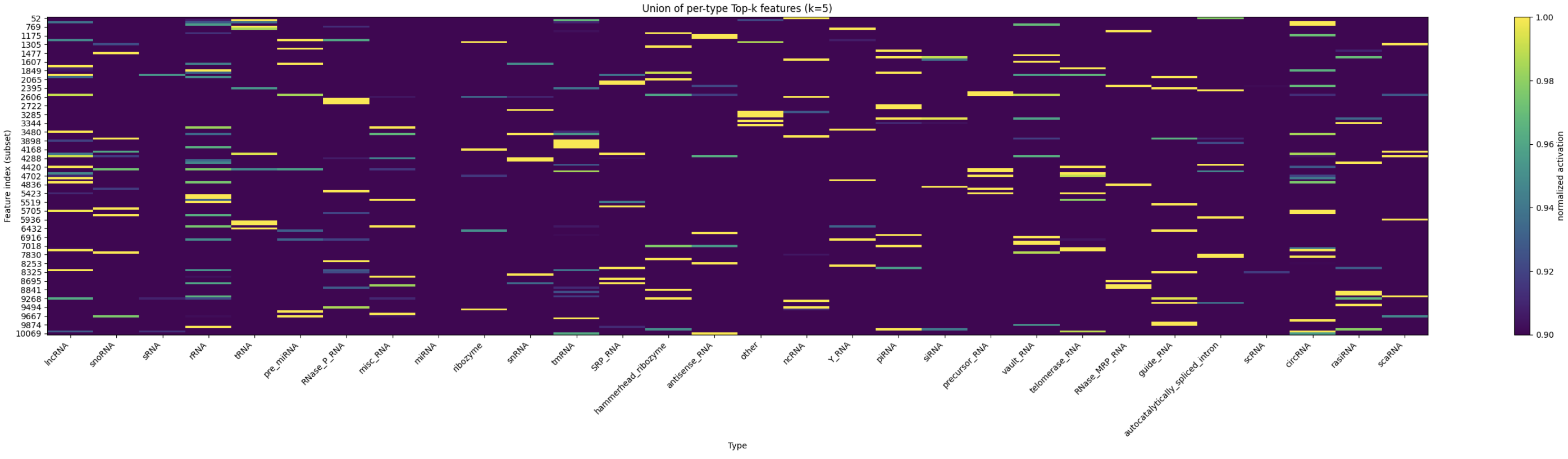}\\[-2mm]
\includegraphics[width=0.7\linewidth]{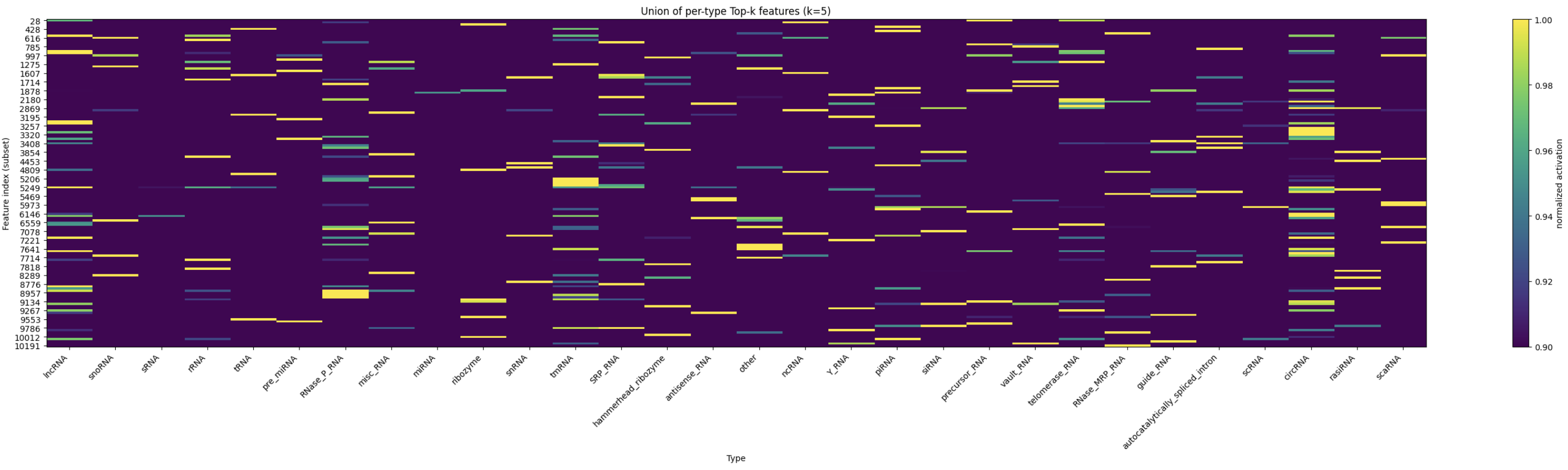}\\[-2mm]
\includegraphics[width=0.7\linewidth]{heatmap_per_type_layer33.png}
\caption{\textbf{Union of per-type top-$k$ features} Color indicates normalized activation; from top to bottom: layer 1, layer 9, layer 18, layer 24, layer 30, and layer 33.}
\label{fig:layer-heatmaps-appendix}
\end{figure}

\clearpage
\section{SAE TRAINING CURVES}\label{app:training-curves}


\begin{figure}[H]
\centering
\includegraphics[width=0.88\linewidth]{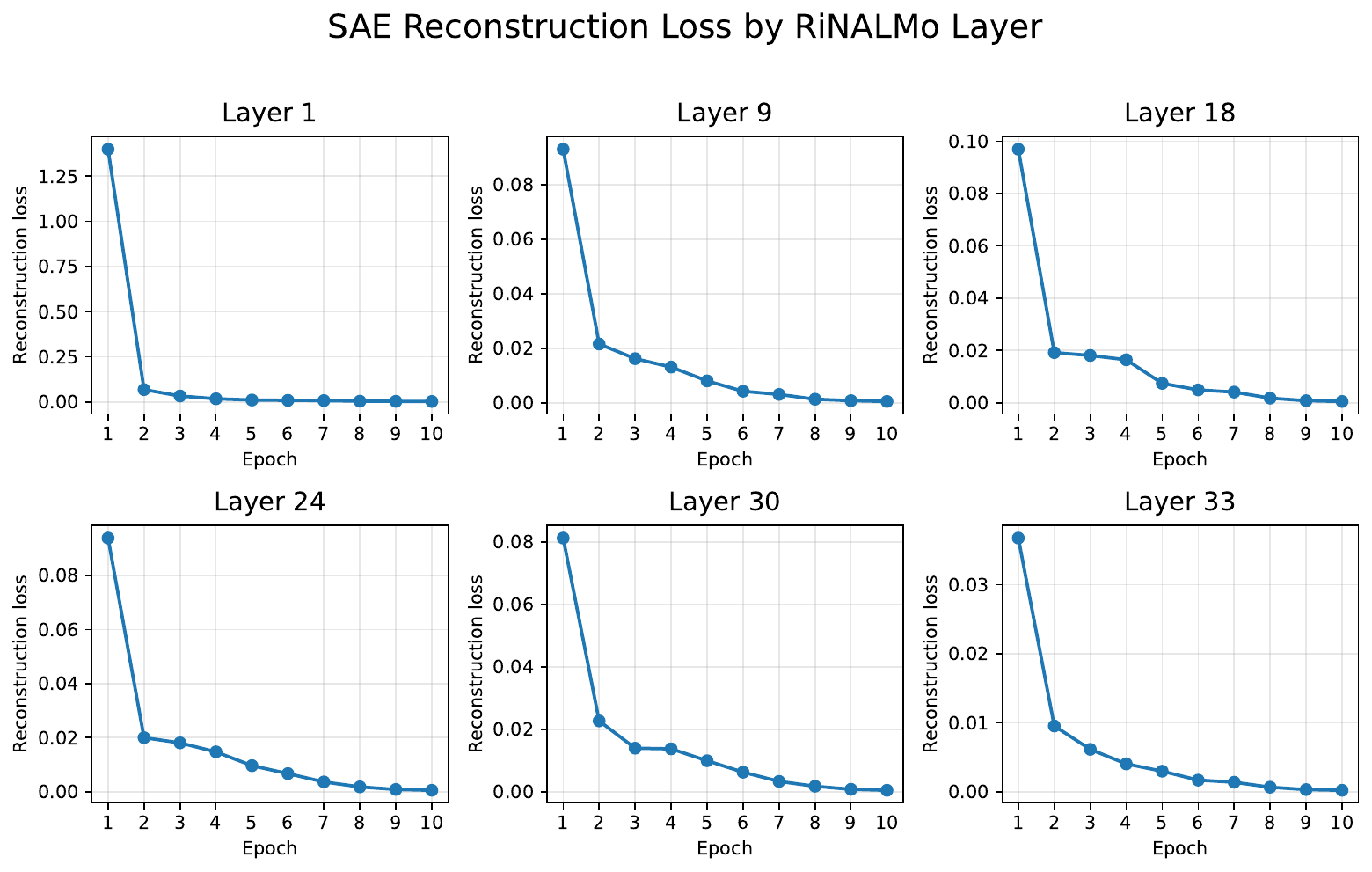}
\caption{Layer-wise SAE reconstruction-loss curves for the RiNALMo layers used in the paper (\(L1, L9, L18, L24, L30, L33\)). Each subplot shows the mean reconstruction error across training epochs for one SAE. For layer \(L1\), the plotted trace corresponds to the final complete 10-epoch run recorded in the provided log file.}
\label{fig:training-curves-appendix}
\end{figure}

\end{document}